\documentclass[conference,comsoc]{IEEEtran}

\usepackage{amsmath,amsthm,amsfonts,braket,algorithm,graphicx,fullpage}

\theoremstyle{definition} 
\theoremstyle{definition}

\newcommand{\kb}[1]{\mathbf{#1}}
\newcommand{\bkm}[2]{\braket{#1|#2}}
\newcommand{\bk}[1]{\braket{#1|#1}}

\newcommand{\prf}[2]{p_{#1,#2}^{A\rightarrow B}}
\newcommand{\prr}[1]{p_{#1}^{A\rightarrow A}}
\newcommand{\prref}[2]{p_{#1,R,#2}^{A\rightarrow A}}

\floatname{algorithm}{Protocol}

\begin{document}
\title{Limited Resource Semi-Quantum Key Distribution}
\author{
\IEEEauthorblockN{Walter O. Krawec}
\IEEEauthorblockA{Computer Science \& Engineering Department\\
University of Connecticut\\
Storrs, CT 06268\\
Email: walter.krawec@uconn.edu}
\and
\IEEEauthorblockN{Eric P. Geiss}
\IEEEauthorblockA{Computer Science Department\\
Iona College\\
New Rochelle, NY 10801}
}

\maketitle

\begin{abstract}
A semi-quantum key distribution (SQKD) protocol allows a quantum user and a limited ``classical'' user to establish a shared secret key secure against an all-powerful adversary.  In this work, we present a new SQKD protocol where the quantum user is also limited in her measurement capabilities.  We describe the protocol, prove its security, and show its noise tolerance is as high as ``fully quantum'' QKD protocols.
\end{abstract}

\section{Introduction}

Quantum Key Distribution (QKD) protocols (see \cite{QKD-survey} for a general survey) allow for the establishment of secret keys between two parties, customarily referred to as Alice ($A$) and Bob ($B$) in a manner secure against even an all-powerful, unbounded adversary Eve ($E$) - i.e., an adversary bounded only by the laws of physics.  Achieving this is impossible through classical communication alone.  QKD protocols require both parties, $A$ and $B$ to be ``quantum'' in that they can both manipulate qubits (or other quantum resources) in certain ways (e.g., both parties must be able to prepare and/or measure qubits in two or more bases).  Semi-quantum Key Distribution (SQKD) protocols, which were first introduced in \cite{SQKD-first} to study the question ``how quantum does a protocol need to be in order to gain an advantage over its classical counterpart?'' do not impose this requirement on both users. In the semi-quantum model, one of the users, typically $B$, is limited to only operating directly in the computational $Z$ basis (spanned by $\ket{0}$ and $\ket{1}$).  The other user, $A$, is ``fully quantum'' in that she can prepare and measure qubits in arbitrary bases.

All SQKD protocols require the use of a two-way quantum channel, allowing a qubit to travel from fully-quantum $A$, to ``classical'' or semi-quantum $B$, then back to $A$.  This also allows an attacker two opportunities to interact with the traveling qubit, thus greatly increasing the complexity of their security analysis.  In fact, though several SQKD protocols have been developed, only recently have unconditional security proofs been derived \cite{SQKD-Krawec-SecurityProof,SQKD-zhang2016single,QKD-krawec2016quantum,SQKD-Multiuser}.  In fact, it was shown in \cite{QKD-krawec2016quantum} that the original SQKD protocol of Boyer et al. from \cite{SQKD-first} has the same tolerance to noise in the quantum channel as the fully quantum BB84 \cite{QKD-BB84} protocol.

\begin{figure}
  \centering
  \includegraphics[width=200pt]{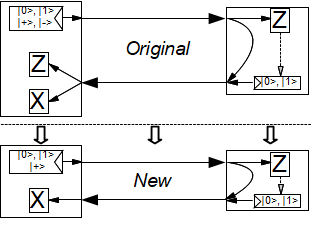}
  \caption{Comparing the original SQKD protocol from \cite{SQKD-first} (top) to the ``limited-resource'' one we analyze here, where Alice's source and measurement capabilities have been curtailed (bottom).}\label{fig:restricted}
\end{figure}

However, all SQKD protocols we are aware of, and certainly all protocols with an unconditional proof of security, require the user $A$ to measure in more than one basis - typically, she must choose to measure in either the $Z$ or the Hadamard $X$ basis each iteration (the latter consisting of states $\ket{\pm} = \frac{1}{\sqrt{2}}(\ket{0}\pm\ket{1})$).  In this paper, we propose a very restricted SQKD protocol where the fully quantum user can only send qubits of the form $\ket{0}$, $\ket{1}$, and $\ket{+}$ (but not $\ket{-}$) and, furthermore, she can only measure in the $X$ basis.  With this protocol, the users are very limited to what channel statistics they may observe.  See Figure \ref{fig:restricted} for a schematic diagram of the capabilities of $A$ and $B$.

We will prove the unconditional security of this new protocol.  In our proof, we will utilize multiple channel statistics, including those gained from mismatched measurements \cite{QKD-Tom-First,QKD-Tom-KeyRateIncrease}.  Interestingly, as we will show, our new, limited, protocol is actually \emph{insecure} without these statistics (to our knowledge, this is the first time mismatched measurements have been used to prove the security of a protocol which is insecure without them - generally they are used only to improve the key-rate bound).  However, by careful use of mismatched statistics, we will show, remarkably, that our protocol has the same noise tolerance as the BB84 protocol, Boyer et al.'s original SQKD protocol \cite{SQKD-first}, and LM05 \cite{QKD-TwoWay-LM05}!

While there is a clear theoretical interest in studying these ``limited resource'' SQKD protocols in order to answer the question ``how quantum'' need a protocol really be, there are also \emph{potential practical benefits to this study}.  For one thing, we show that fewer measurement devices (which are expensive) are required.  Secondly, one could envision scenarios where equipment potentially breaks down while in service.  If the equipment is installed, say, on a satellite, it would not only be expensive to repair or replace the damaged system, but also it would take time to launch a mission to do so.  Our work here shows how fully-quantum systems, when endowed with the possibility of two-way quantum communication, could still be used (as a ``semi-quantum'' system) should certain measurement devices fail.


\subsection{Notation}
We assume a general understanding of quantum communication and information theory; for more information on these, the reader is referred to \cite{QC-intro}.  In this section, we will describe the notation used throughout the paper.

We denote by $H(p_1, \cdots, p_n) = -\sum_ip_i\log p_i$ to be the Shannon entropy (all logarithms in this paper are base $2$ unless otherwise noted).  If $n=2$, then we simply write $h(p_1) = h(p_2)$, the binary entropy function.    If $\rho$ is a density operator (a positive semi-definite Hermitian operator of unit trace) acting on some finite dimensional Hilbert space $\mathcal{H}$ (in this case $\rho$ models a quantum system), then we write $S(\rho)$ to mean the von Neumann entropy of $\rho$ which is simply $S(\rho) = H(\lambda_1, \cdots, \lambda_m)$ where the $\lambda_i$'s are the eigenvalues of $\rho$.  We write $\rho_{AB}$ to mean a density operator acting on some joint Hilbert space $\mathcal{H}_A\otimes\mathcal{H}_B$ and $\rho_B$ to mean the partial trace over $\mathcal{H}_A$ (i.e., $\rho_B = tr_A\rho_{AB}$ is an operator acting on $\mathcal{H}_B$).  We use $S(A|B)_\rho$ to mean the conditional von Neumann entropy of $A$'s system conditioned on $B$'s, namely: $S(A|B)_\rho = S(AB)_\rho - S(B)_\rho = S(\rho_{AB}) - S(\rho_B)$.  Often we will forgo writing the subscript ``$\rho$'' when the context is clear.

The computational $Z$ basis are states $\{\ket{0}, \ket{1}\}$ and the Hadamard $X$ basis are denoted by states $\{\ket{+}, \ket{-}\}$, where $\ket{\pm} = \frac{1}{\sqrt{2}}(\ket{0}\pm\ket{1})$.  Given a vector $\ket{v}$, we write $\kb{v}$ to mean $\ket{v}\bra{v} = vv^*$.  Similarly, $\kb{0} = \ket{0}\bra{0}$ and $\kb{1} = \ket{1}\bra{1}$, and so on.  Also, we write $\kb{0}_A$ to mean $\ket{0}\bra{0}$ acting on $A$'s Hilbert space (and similarly for $\kb{i}_A$ and $\kb{i}_B$).

\section{The Protocol}

The protocol we analyze is semi-quantum and, so, $B$ being the ``classical'' user, is able only to work directly with the computational $Z$ basis.  Namely, $B$ on receiving a qubit from $A$, is allowed to perform one of two operations:
1: he may \emph{Measure and Resend}: that is, he will subject the incoming qubit to a measurement in the $Z = \{\ket{0}, \ket{1}\}$ basis.  If this measurement results in outcome $\ket{r}$, for $r \in \{0,1\}$, he will send the qubit $\ket{r}$ back to $A$;
or 2: he may \emph{Reflect}: that is, he will ignore the incoming qubit, and ``bounce'' or reflect it back to $A$.  In this case, $B$ does not disturb the state of the qubit, but also does not learn anything about its state (essentially $A$ is talking to herself in this case).

The other user, $A$, is allowed to prepare arbitrary qubits and to perform arbitrary qubit measurements.  However, in this paper, unlike prior semi-quantum protocols, we will place a further restriction on $A$ in that she is only able to measure in the $X$ basis.  We do, however, allow $A$ to send qubits in either basis, however we will only require her to send three states: $\ket{0}$, $\ket{1}$, or $\ket{+}$.  In a way, this protocol may be considered a limited version of the three state SQKD protocol described in \cite{SQKD-lessthan4}; here, however, we require that $A$ measure only in the $X$ basis (the protocol in \cite{SQKD-lessthan4} permitted her to measure in both bases).  The protocol is described in Protocol \ref{prot:sqkd}.

\begin{algorithm}
\caption{Limited-SQKD}\label{prot:sqkd}

\textbf{Quantum Communication Stage}:
The quantum communication stage of the protocol repeats the following process:

1.) With probability $p/2$ $A$ sends $\ket{0}$; with probability $p/2$ she sends $\ket{1}$, with probability $1-p$, she sends $\ket{+}$

2.) $B$ chooses, with probability $q$, to measure and resend or, with probability $1-q$ to reflect the qubit.  If he chooses the former, he saves his measurement result to potentially serve as his \emph{raw key} bit for this iteration.

3.) $A$ measures the returning qubit in the $X$ basis.

4.) Using the authenticated classical channel, $A$ divulges her initial choice of basis and $B$ his choice of operation.

5.) If $A$ chose to send in the $Z$ basis and if $B$ chose to measure and resend, they may use this iteration to contribute to their raw key ($A$ using her initial preparation choice and $B$ his measurement outcome).  All other iterations, along with a suitably sized, randomly chosen subset of iterations of the former kind, may be used to estimate the channel statistics.  Note that we will not discard iterations: indeed, as we will see, the use of \emph{mismatched measurements} \cite{QKD-Tom-First,QKD-Tom-KeyRateIncrease} will be vital to proving the security of this limited SQKD protocol and without them, the protocol is in fact insecure.
\end{algorithm}

\subsection{Security of QKD Protocols}

We first consider security against \emph{collective attacks} \cite{QKD-survey} whereby $E$ attacks each signal independently and identically but is allowed to postpone her joint quantum measurement until later.  We will later consider security against general attacks (thus unconditional security).

Let $N$ be the size of the raw key after $A$ and $B$ perform the quantum communication stage of an (S)QKD protocol.  This is a classical string that is partially correlated and partially secret.  $A$ and $B$ will perform an error correction protocol (leaking additional information to $E$) followed by a privacy amplification protocol.  The end result is a secret key of size $\ell(N)$.  It was shown in \cite{QKD-winter-keyrate} that the \emph{key rate}, denoted $r$, of such a process in the \emph{asymptotic scenario is}:
$r := \lim_{N\rightarrow \infty}\frac{\ell(N)}{N} = \inf[S(A|E) - H(A|B)],$ where the infimum is over all collective attacks which induce the observed statistics.
Our goal, then, is to determine a lower-bound on our protocol's key rate \emph{as a function only on observed parameters}.  We are particularly interested in determining for what noise levels is $r$ positive - i.e., we wish to determine how noisy the quantum channel can be before the key rate drops to zero.

\subsection{On the Need for Mismatched Measurements}

In this subsection we show Protocol \ref{prot:sqkd} is actually \emph{insecure} if $A$ and $B$ limit their parameter estimation only to considering ``error events'' (e.g., a $\ket{+}$ flipping to a $\ket{-}$ when $B$ reflects).  We show a very simple attack $E$ may perform, which induces no noise, yet causes her to gain full information on the raw key.  This attack may be detected by considering mismatched measurement events  \cite{QKD-Tom-First,QKD-Tom-KeyRateIncrease} (e.g., a $\ket{0}$ being measured by $A$ as a $\ket{+}$).

The attack strategy proceeds as follows: in the forward channel, $E$ will ignore the qubit - thus there will be no $Z$ basis noise (we make the usual assumption that any noise is the result of the attacker \cite{QKD-survey}).  In the reverse channel, $E$ will apply a unitary operator $\mathcal{U}_R$ acting on the qubit and $E$'s private memory as follows:
\begin{align*}
\mathcal{U}_R \ket{+,0} &= \ket{+,0} && \mathcal{U}_R \ket{-,0} = \ket{+,1},
\end{align*}
where the first subspace is the qubit and the second is $E$'s quantum ancilla which we assume is initially cleared to some ``zero'' state $\ket{0}$ (this is a collective attack).

It is clear that the above is a unitary operator.  Furthermore, it induces no $X$ basis noise - since $A$ cannot send $\ket{-}$, this attack goes undetected.  However, by linearity of $\mathcal{U}_R$, we have:
\begin{align*}
\mathcal{U}_R \ket{0,0} &= \ket{+,+}&&\mathcal{U}_R \ket{1,0} = \ket{+,-}.
\end{align*}
Thus, if $B$ measures and resends, a simple measurement of $E$'s ancilla in the $X$ basis provides her with full information on the raw key bit of $B$ (and thus also $A$).  This attack also goes undetected.

One could foil this attack by having $A$ send also a $\ket{-}$ in which case the above attack would induce an $X$ basis error rate of $50\%$.  However, we may also foil the above attack, without altering the quantum portion of our protocol (i.e., without increasing the quantum complexity of the protocol), by considering mismatched measurements.

Indeed, the idea is as follows: if there were no attack in the reverse channel, one would expect $A$ to observe $\ket{+}$ or $\ket{-}$ with equal probability in the event $B$ chose to measure and resend.  $\mathcal{U}_R$, however, causes $A$ to always measure $\ket{+}$ and so would be detected.

Obviously, $\mathcal{U}_R$ is just one explicit attack which we constructed to illustrate the sensitivity of this new protocol.  In the following, we will derive a lower-bound on the key rate of our protocol, applicable for any attack.

\subsection{Proof of Security}

Our goal now is to determine a bound on the key rate of Protocol \ref{prot:sqkd} based only on statistics that may be observed directly by $A$ and $B$.  To do so, we must first describe the joint quantum system held by $A$ and $B$ in the event a particular iteration was used to contribute towards their raw key.  In particular, in the event $A$ sends a $Z$ basis state and $B$ measures and resends.  We first consider \emph{collective attacks}; later, we will discuss security against arbitrary, \emph{general attacks}.

In the following we will denote by $\mathcal{H}_T$ to be the two-dimensional Hilbert space modeling the traveling qubit (i.e., it is the ``transit'' space) while $\mathcal{H}_E$ is the, without loss of generality, finite-dimensional Hilbert space modeling $E$'s quantum memory.  As we are considering, for the time being, collective attacks, we may describe $E$'s collective attack as a pair of unitary operators $(U_F, U_R)$ where $U_F$ is applied in the forward direction (as the qubit travels from $A$ to $B$) and where $U_R$ is applied in the reverse (as the qubit travels from $B$ back to $A$).  Furthermore, we may assume that $E$'s ancilla is cleared to some pure ``zero'' state $\ket{0}_E$.  Therefore, we may write the action of $E$'s attack operators as follows:
$U_F\ket{0,0}_{TE} = \ket{0, e_0} + \ket{1, e_1}$;
$U_F\ket{1,0}_{TE} = \ket{0, e_2} + \ket{1, e_3}$; and
$U_R\ket{i, e_j}_{TE} = \ket{0, e_{i,j}^0} + \ket{1, e_{i,j}^1};$
where the various $\ket{e_i}$ and $\ket{e_{i,j}^k}$ are arbitrary, not necessarily normalized nor orthogonal, states in $E$'s ancilla $\mathcal{H}_E$.  Unitarity of $U_F$ and $U_R$ impose various restrictions on these states which will become important later.

With the above notation described, we may now derive the joint quantum system modeling a single iteration of the protocol, conditioning on events leading to a key bit being distilled (so as to compute the key rate equation).  Conditioning on the event this iteration is used to contribute towards the raw key, it holds that $A$ sends $\ket{0}$ or $\ket{1}$ with probability $1/2$ each.  $E$ subsequently attacks the qubit using $U_F$, and then forwards the qubit to $B$ who measures it in the $Z$ basis, resending his measurement result as a new qubit.  This qubit is again captured by $E$ who attacks with $U_R$.  Finally, $A$ performs an $X$ basis measurement, discarding the result (effectively tracing out $\mathcal{H}_T$). Thus, the desired system, denoted $\rho_{ABE}$, is:
\begin{align*}
&\frac{1}{2}\kb{0}_A\kb{0}_B\otimes(\kb{e_{0,0}^0} + \kb{e_{0,0}^1}) + \frac{1}{2}\kb{0}_A\kb{1}_B\otimes(\kb{e_{1,1}^0} + \kb{e_{1,1}^1})\\
+&\frac{1}{2}\kb{1}_A\kb{0}_B\otimes(\kb{e_{0,2}^0} + \kb{e_{0,2}^1}) + \frac{1}{2}\kb{1}_A\kb{1}_B \otimes (\kb{e_{1,3}^0} + \kb{e_{1,3}^1}).
\end{align*}
We now use Theorem 1 of \cite{QKD-krawec2016quantum} to bound $S(A|E)$ as follows:
\begin{align}
S(A|E) & \ge \frac{\bk{e_{0,0}^0} + \bk{e_{1,3}^1}}{2} \cdot\Delta_1\label{eq:keyrate-bound}\\
&+ \frac{\bk{e_{0,0}^1} + \bk{e_{1,3}^0}}{2} \cdot\Delta_2\notag\\
&+ \frac{\bk{e_{1,1}^1} + \bk{e_{0,2}^0}}{2} \cdot\Delta_3\notag\\
&+ \frac{\bk{e_{1,1}^0} + \bk{e_{0,2}^1}}{2} \cdot\Delta_4\notag
\end{align}
where:
\begin{align}
\Delta_1 &= h\left( \frac{\bk{e_{0,0}^0}}{\bk{e_{0,0}^0} + \bk{e_{1,3}^1}}\right) - h(\lambda_1)\\
\Delta_2 &= h\left( \frac{\bk{e_{0,0}^1}}{\bk{e_{0,0}^1} + \bk{e_{1,3}^0}}\right) - h(\lambda_2)\\
\Delta_3 &= h\left( \frac{\bk{e_{1,1}^1}}{\bk{e_{1,1}^1} + \bk{e_{0,2}^0}}\right) - h(\lambda_3)\\
\Delta_4 &= h\left( \frac{\bk{e_{1,1}^0}}{\bk{e_{1,1}^0} + \bk{e_{0,2}^1}}\right) - h(\lambda_4)
\end{align}
and:
\begin{align}
\lambda_1 &= \frac{1}{2} + \frac{\sqrt{(\bk{e_{0,0}^0} - \bk{e_{1,3}^1})^2 + 4Re^2\bkm{e_{0,0}^0}{e_{1,3}^1}}}{2(\bk{e_{0,0}^0} + \bk{e_{1,3}^1})}\label{eq:l1}\\
\lambda_2 &= \frac{1}{2} + \frac{\sqrt{(\bk{e_{0,0}^1} - \bk{e_{1,3}^0})^2 + 4Re^2\bkm{e_{0,0}^1}{e_{1,3}^0}}}{2(\bk{e_{0,0}^1} + \bk{e_{1,3}^0})}\label{eq:l2}\\
\lambda_3 &= \frac{1}{2} + \frac{\sqrt{(\bk{e_{1,1}^1} - \bk{e_{0,2}^0})^2 + 4Re^2\bkm{e_{1,1}^1}{e_{0,2}^0}}}{2(\bk{e_{1,1}^1} + \bk{e_{0,2}^0})}\label{eq:l3}\\
\lambda_4 &= \frac{1}{2} + \frac{\sqrt{(\bk{e_{1,1}^0} - \bk{e_{0,2}^1})^2 + 4Re^2\bkm{e_{1,1}^0}{e_{0,2}^1}}}{2(\bk{e_{1,1}^0} + \bk{e_{0,2}^1})}\label{eq:l4}
\end{align}

We now show how parameter estimation may be used to determine bounds on the various inner-products needed to evaluate this lower-bound on $S(A|E)$.  We will also take advantage of information learned through mismatched measurements (which, as discussed earlier, is actually critical to the security of this protocol).  In \cite{QKD-krawec2016quantum}, these statistics were used to evaluate the key rate of Boyer et al.'s \cite{SQKD-first} original SQKD protocol.  We will extend the results from that paper and apply them to our new protocol.  The difficulty is that $A$ is not allowed to measure in the $Z$ basis, thus limiting the information we can learn (unlike in \cite{QKD-krawec2016quantum} where $A$ was allowed to measure in any basis of her choice).

Denote by $\prf{i}{j}$ to be the probability that if $A$ sends $\ket{i}$ (for $i \in \{0, 1, +\}$) then $B$ measures $\ket{j}$ (for $j \in \{0,1\}$).  It is obvious that $A$ and $B$ may observe these probabilities for all combinations of $i$ and $j$.  In particular, we clearly have: $\prf{0}{0} = \braket{e_0|e_0} = \bk{e_{0,0}^0} + \bk{e_{0,0}^1}$ (the last equality follows from unitarity of $U_R$).  Since each $\bk{e_{i,j}^k}\ge 0$, this gives us a bound on these inner-products.  Similar restrictions for the other $\bk{e_{i,j}^k}$ may be found based on the observable quantity $\prf{a}{b}$.

What remains to be shown is how to determine bounds on those quantities appearing inside the $\lambda_i$ expressions.  Denote by $\prr{i,j,k}$ to be the probability that $A$ measures $\ket{k}$ conditioning on the event $A$ initially sent $\ket{i}$ and $B$ measured (and thus resent) outcome $\ket{j}$.  Here we have $i \in \{0,1,+\}$, $j \in \{0,1\}$ and $k \in \{+, -\}$.  Also, denote by $\prref{i}{k}$ to be the probability that, if $A$ initially sends $\ket{i}$, and if $B$ reflects, than $A$ measures $\ket{k}$.

Consider, first, the quantity $Q_X = \prref{+}{-}$: the probability of a $\ket{+}$ flipping to a $\ket{-}$ if $B$ reflects (note that the users may not measure $\prref{-}{+}$ as $A$ is not allowed to send the state $\ket{-}$).  In the event $B$ reflects, his operation is, essentially, the identity operator.  Thus, we may think of $E$'s attack as a single operator $V = U_RU_F$; that is, $A$ sends a qubit, $E$ attacks with $V$, and $A$ measures a qubit.  Recalling that $E$'s memory is cleared to some ``zero'' state at the start of each iteration, let us denote $V$'s action as follows:
$V\ket{0,0}_{TE} = \ket{0, g_0} + \ket{1, g_1}$; and 
$V\ket{1,0}_{TE} = \ket{0,g_2} + \ket{1, g_3};$
where, due to linearity of the operators $U_F$ and $U_R$, we find:
\begin{align}
\ket{g_0} = \ket{e_{0,0}^0} + \ket{e_{1,1}^0} && \ket{g_1} = \ket{e_{0,0}^1} + \ket{e_{1,1}^1}\label{eq:V-action}\\
\ket{g_2} = \ket{e_{0,2}^0} + \ket{e_{1,3}^0}&&\ket{g_3} = \ket{e_{0,2}^1} + \ket{e_{1,3}^1}\notag
\end{align}

Due to linearity of $V$, we easily find:
\begin{align}
Q_X &=  \frac{1}{2} - \frac{1}{2}Re(\bkm{g_0}{g_1} + \bkm{g_0}{g_3} + \bkm{g_1}{g_2} + \bkm{g_2}{g_3}).\label{eq:QX-first}
\end{align}
(Note that, above, we made use of the fact that $\bkm{g_0}{g_2} + \bkm{g_1}{g_3} = 0$ due to unitarity of $V$.)

Let us first consider $\bkm{g_0}{g_1}$ and $\bkm{g_2}{g_3}$.  As demonstrated in \cite{QKD-krawec2016quantum}, the statistics $\prref{0}{+}$ and $\prref{1}{+}$ can be used to determine these quantities.  In particular:
\begin{align}
&\prref{0}{+} = \frac{1}{2}(\bk{g_0} + \bk{g_1} + 2Re\braket{g_0|g_1})\notag\\
\Rightarrow &Re\bkm{g_0}{g_1} = \prref{0}{+} - \frac{1}{2},\label{eq:reg01}
\end{align}
where, for the second derivation, we used the fact that $\bk{g_0} + \bk{g_1} = 1$, due to unitarity of $V$.  Furthermore, note that if $E$'s attack is symmetric in that $\prref{0}{+} = 1/2$, then $Re\bkm{g_0}{g_1} = 0$.  However, we do not require this symmetry assumption.

Similarly, we may derive the following:
\begin{equation}\label{eq:reg23}
Re\bkm{g_2}{g_3} = \prref{1}{+} - \frac{1}{2}.
\end{equation}


Let us now turn our gaze to the remaining two quantities $Re\bkm{g_0}{g_3}$ and $Re\bkm{g_1}{g_2}$, from which we will acquire bounds on the desired quantities appearing in the $\lambda_i$.  From Equation \ref{eq:V-action}, we may expand these ``$g$'' states in terms of ``$e$'' states:

\begin{align*}
\bkm{g_0}{g_3} &= \overbrace{\bkm{e_{0,0}^0}{e_{1,3}^1}}^{\Lambda_1} + \bkm{e_{0,0}^0}{e_{0,2}^1} + \overbrace{\bkm{e_{1,1}^0}{e_{0,2}^1}}^{\Lambda_4} + \bkm{e_{1,1}^0}{e_{1,3}^1}\\
\bkm{g_1}{g_2} &= \underbrace{\bkm{e_{0,0}^1}{e_{1,3}^0}}_{\Lambda_2} + \bkm{e_{0,0}^1}{e_{0,2}^0} + \underbrace{\bkm{e_{1,1}^1}{e_{0,2}^0}}_{\Lambda_3} + \bkm{e_{1,1}^1}{e_{1,3}^0}
\end{align*}
where, above, we defined $\Lambda_i$ to be that inner product required to compute $\lambda_i$.

From \cite{QKD-krawec2016quantum}, we know that:

\begin{align}
\eta_1 &= Re\braket{e_{0,0}^0|e_{0,2}^1} + \braket{e_{0,0}^1|e_{0,2}^0} \label{eq:sqkd-sum1}\\
&=2\prf{+}{0}\prr{+,0,+} - \frac{1}{2}(\prf{0}{0}+\prf{1}{0})\notag\\
&- \prf{0}{0}\left(\prr{0,0,+}-\frac{1}{2}\right) - \prf{1}{0}\left(\prr{1,0,+}-\frac{1}{2}\right)\notag\\
& - \prf{+}{0} +\frac{1}{2}(\prf{0}{0}+\prf{1}{0}).\notag
\end{align}
\begin{align}
\eta_2 &= Re\braket{e_{1,1}^0|e_{1,3}^1} + \braket{e_{1,1}^1|e_{1,3}^0}\label{eq:sqkd-sum2}\\
&=2\prf{+}{1}\prr{+,1,+} - \frac{1}{2}(\prf{0}{1}+\prf{1}{1})\notag\\
&- \prf{0}{1}\left(\prr{0,1,+}-\frac{1}{2}\right) - \prf{1}{1}\left(\prr{1,1,+}-\frac{1}{2}\right)\notag\\
& + \prf{+}{0} -\frac{1}{2}(\prf{0}{0}+\prf{1}{0}).\notag
\end{align}

The above expressions are easily discovered by tracing the evolution of a qubit $\ket{+}$ as it is attacked by $E$, then measured by $B$ (conditioning on a particular outcome), attacked by $E$ again, and finally measured in the $X$ basis by $A$.  Even though the protocol considered in \cite{QKD-krawec2016quantum}, from which the above identities were derived, did not involve a limited $A$, these quantities are all based on statistics which may be observed in our limited case as none require $A$ to measure in a non-$X$ basis.  

At this point, we already have enough information to numerically evaluate our lower bound with surprisingly good results.  Indeed, the (eight) variables we optimize over are:
\[
\bk{e_{0,0}^1}, \bk{e_{1,3}^1}, \bk{e_{0,2}^1}, \bk{e_{1,1}^1}, \{\Lambda_i\}_{i=1}^4
\]
subject to the following restrictions (also listed are the reasons for the given restrictions):
\begin{align*}
&\textbf{Restriction} && \textbf{ Reason}\\
\bk{e_{0,0}^0} &= \prf{0}{0} - \bk{e_{0,0}^1} && \text{Unitarity of $U_R$}\\
\bk{e_{1,3}^0} &= \prf{1}{1} - \bk{e_{1,3}^1} && \text{Unitarity of $U_R$}\\
\bk{e_{0,2}^0} &= \prf{1}{0} - \bk{e_{0,2}^1} && \text{Unitarity of $U_R$}\\
\bk{e_{1,1}^0} &= \prf{0}{1} - \bk{e_{1,1}^1} && \text{Unitarity of $U_R$}\\
|\Lambda_1| &\le \sqrt{\bk{e_{0,0}^0}\bk{e_{1,3}^1}} && \text{Cauchy Schwarz}\\
|\Lambda_2| &\le \sqrt{\bk{e_{0,0}^1}\bk{e_{1,3}^0}} && \text{Cauchy Schwarz}\\
|\Lambda_3| &\le \sqrt{\bk{e_{1,1}^1}\bk{e_{0,2}^0}} && \text{Cauchy Schwarz}\\
|\Lambda_4| &\le \sqrt{\bk{e_{1,1}^0}\bk{e_{0,2}^1}} && \text{Cauchy Schwarz}
\end{align*}
And, finally, one further restriction required by Equation \ref{eq:QX-first} combined with Equations \ref{eq:reg01}, \ref{eq:reg23}, \ref{eq:sqkd-sum1}, and \ref{eq:sqkd-sum2}:
\begin{align}
Q_X &= 1 - \frac{1}{2}(\Lambda_1+\Lambda_2 + \Lambda_3 + \Lambda_4 + \eta_1 + \eta_2)\label{eq:QX-bound}\\
&-\frac{1}{2}(\prref{0}{+} + \prref{1}{+})\notag
\end{align}

We numerically optimize Equation \ref{eq:keyrate-bound} subject to the above restrictions for various levels of noise.  We consider two common scenarios to illustrate our bound (though, our bound is applicable to \emph{any two-way qubit channel}): first, the two quantum channels are modeled as independent depolarization channels whereby $Q_X = 2Q(1-Q)$ and $Q$ is the $Z$ basis error rate in the forward (and, independently, the reverse) channel; secondly, the two channels are modeled as dependent channels whereby $Q_X = Q$.  In the first case, we attain a noise tolerance of $7.9\%$ (i.e., the key rate expression is positive for all $Q \le 7.9\%$).  In the second, we attain a noise tolerance of $11\%$.  The tolerance in both these scenarios is exactly the same as the original Boyer et al., SQKD protocol without restrictions on $A$!  This is also the same noise tolerance that BB84 could suffer with these noise parameters \cite{QKD-renner-keyrate} and also the fully-quantum LM05 QKD protocol \cite{QKD-twoway2}.  The user $A$ may therefore utilize a simpler device compared to the original Boyer et al., protocol; alternatively, if $A$ is equipped with multiple measurement devices, and the $Z$ measurement device fails during operation, it may still be possible to do QKD between the two users.  

Note that the above analysis assumed collective attacks.  However, this protocol may be made permutation invariant in the usual way (e.g., by publicly choosing a random permutation and permuting the raw key).  In this case, the results from \cite{QKD-general-attack,QKD-general-attack2} apply and we attain security against general attacks (\emph{and thus unconditional security}).  Indeed, we may perform the above security analysis under an equivalent entanglement based version of the protocol whereby $A$ prepares a maximally entangled state and $B$'s measurement operation is modeled as a CNOT operation.

\section{Closing Remarks}

We have taken the semi-quantum protocol of Boyer et al. \cite{SQKD-first}, removed $A$'s ability to measure in any basis other than the $X$-basis, proved its security, and have shown that, despite this limitation, it can suffer the same amount of noise as the original SQKD protocol and also ``fully quantum'' protocols.  An obvious question for future work would be: can we restrict $A$'s capabilities even further (where $A$ sends fewer than three states)?  It would also be interesting to analyze (and compare) the key rate in the finite key setting.

\section*{Acknowledgment}
This work was partially completed while WK was at Iona College, New Rochelle NY, USA.


\end{document}